\documentclass[seceq]{ptptex}

\title{Clebsch Potentials in the Variational Principle for a Perfect Fluid}

\author{Hiroki \textsc{Fukagawa}$^{1}$%
\thanks{E-mail: hiroki@beer.appi.keio.ac.jp%
}and Youhei \textsc{Fujitani}$^{1,2}$%
\thanks{E-mail: youhei@appi.keio.ac.jp%
} }

\inst{$^{1}$School of Fundamental Science \& Technology, Keio University,\\
Yokohama 223-8522, Japan\\
 $^{2}$Institute of Industrial Science, University of Tokyo,\\
Tokyo 153-8505, Japan}

\abst{Equations for a perfect fluid can be obtained by means of the variational
principle both in the Lagrangian description and in the Eulerian one.
It is known that we need additional fields somehow to describe a rotational
isentropic flow in the latter description. We give a simple explanation
for these fields; they are introduced to fix both ends of a pathline
in the variational calculus. This restriction is imposed in the former
description, and should be imposed in the latter description. It is
also shown that we can derive a canonical Hamiltonian formulation
for a perfect fluid by regarding the velocity field as the input in
the framework of control theory.}

\begin{document} 
\maketitle

\section{Introduction\label{sec:Introduction}}

The Euler equation, together with mass and entropy conservations,
describes dynamics of the perfect fluid, and can be derived from the
variational principle in the Lagrangian description \cite{Bennett:}.
Let us write $\rho$ for mass per unit \emph{volume}, $s$ for entropy
per unit \emph{mass}, and $\boldsymbol{v}$ for velocity field. Because
of the local equilibrium, the internal-energy density per unit \emph{mass},
$\epsilon$, is a function of $\rho\ {\rm and}\ s$. Apart from constraints
coming from the conservation laws, the Lagrangian density is given
by \begin{equation}
\rho\left\{ \frac{1}{2}\boldsymbol{v}^{2}-\epsilon(\rho,s)\right\} \ .\label{eqn:intro}\end{equation}
 The action, which is minimized to yield the Euler equation, is given
by the integral of the Lagrangian density over the space and time
considered. In the Lagrangian description, the position should be
regarded as a variable of the Lagrangian density, and thus the velocity
field $\boldsymbol{v}$ is given by the time derivative of the position
of a fluid particle.

In the Eulerian description, not the position but the velocity is
a variable of the Lagrangian density. Minimizing the action, we can
obtain the Euler equation. However, unlike in the Lagrangian description,
the resultant velocity field cannot be rotational on the isentropic
condition. To remove this flaw, Bateman \cite{Bateman} added some
scalar fields, sometimes called Clebsch potentials \cite{Clebsch,lamb},
to the Lagrangian density. Later Lin \cite{Lin} used more additional
scalar fields, which Selinger and Whitham \cite{SELINGER} considered
to be redundant. Schutz \cite{Schutz} also used Bateman's scalar
fields in considering the general-relativistic gravitation field,
and still complained of too many additional scalar fields. Irrespective
of these controversies, the variational principles \cite{Bateman,Lin,Schutz}
in the Eulerian description have been used in some studies on magnetohydrodynamics,
multivalued plasmas, elasticity and a hydrodynamic description of
relativistic stars \cite{SELINGER,Elze,Friedman}. Recently, Kambe
\cite{Kambe} has explained Clebsch potentials from the point of view
of symmetry of the gauge theory. Yoshida \cite{Yoshida,Y2} claimed
that the number of required additional fields should be more than
Bateman's and fewer than Lin's. Some authors \cite{Bennett:,Lin,SELINGER,Kambe,Yoshida,Y2,Holm,cendra,holm2}
considered that Clebsch potentials are related with the Lagrangian
coordinates.

We show below that Clebsch potentials in the variational calculus
of the Euler description are introduced to fix both ends of a pathline;
the same restriction is imposed in that of the Lagrangian description.
In relation to the variational principles, various canonical and noncanonical
Hamiltonian formulations have been proposed \cite{Yoshida,Y2,Holm,cendra,holm2,Arnold,Morrison,Salmon}.
We derive a canonical Hamiltonian formulation by means of an optimal
control theory known as Pontryagin's minimum principle.

We state the problem in more detail and show our notation in \S\ref{sec:Statement-of-the}.
Our explanation for Clebsch potential is given in \S\ref{sub:Fixing-the-endpoints}.
We give a brief review of the control theory in \S\ref{subsec:pont},
and derive a canonical formulation by means of the control theory
in \S\ref{sub:Pontryagin's-minimum-principle}. The last section
is devoted to discussion. We compare our study with previous works
on Clebsch potentials and Hamiltonian formulations in \S\ref{sec:Discussion}.
Some details are relegated to Appendices. Although we limit the following
discussion in the text to nonrelativistic perfect fluid, we show in
Appendix \ref{sec:A-variational-principle} that our discussion is
also valid for a relativistic perfect fluid.

\section{Statement of the problem\label{sec:Statement-of-the}}

We write $V$ for the spacial region of the container filled with
a perfect fluid, and \emph{$\partial V$} for the surface of the container.
We fix the container and consider the dynamics of the fluid from the
initial time $t_{{\rm init}}$ to the final time $t_{{\rm fin}}$.
Let $\tau$ denote time in the Lagrangian description and $t$ denote
time in the Eulerian one although they are equivalent in the nonrelativistic
theory. The partial derivative with respect to $\tau$~($\partial_{\tau}$),
and $t$~($\partial_{t}$), imply the Lagrangian and Eulerian time
derivatives, respectively. See Appendix \ref{sec:The-Lagrangian-time}
for their relation. We label a fluid particle with its initial position
$\boldsymbol{a}$, and write $\boldsymbol{X}(\boldsymbol{a},\tau)$
for its position at time $\tau$. Thus, $\boldsymbol{a}=\boldsymbol{X}(\boldsymbol{a},t_{{\rm init}})$
gives the Lagrangian coordinates. The volume element in the Lagrangian
coordinates can be given by the determinant of the Jacobian matrix,
\begin{equation}
J(\boldsymbol{a},\tau)\equiv\frac{\partial(X_{1},X_{2},X_{3})}{\partial(a_{1},a_{2},a_{3})}\ ,\label{eq:jacobian}\end{equation}
 where $X_{i}$ and $a_{i}$ are, respectively, the components of
$\boldsymbol{X}$ and $\boldsymbol{a}$. By definition, we have $J(\boldsymbol{a},t_{{\rm init}})=1$.
We assume that a fluid particle never shrinks to a point, i.e., $J(\boldsymbol{a},\tau)$
has no singular points, in the space and time considered. Thus, we
can define the inverse of $\boldsymbol{x}=\boldsymbol{X}(\boldsymbol{a},\tau)$,
for which we write $\boldsymbol{a}(\boldsymbol{x},t)$. Let us write
$T$ for temperature and $p$ for pressure, and the first law of thermodynamics,
$d\epsilon=-pd\rho^{-1}+Tds$, yields \begin{equation}
p\equiv\rho^{2}\left(\frac{\partial\epsilon}{\partial\rho}\right)_{s}\ {\rm and\ }T\equiv\left(\frac{\partial\epsilon}{\partial s}\right)_{\rho}\ ,\label{eq:the first_low_of_thermodyanmics}\end{equation}
 where the subscripts $_{s}$ and $_{\rho}$ indicate variables fixed
in the respective partial differentiations. Below, Roman indices run
from 1 to 3 except in $\S$\ref{sec:Hamiltonian-formulations-for}
and Appendix \ref{sub:Poisson-bracket}, and repeated indices are
summed up, unless specified otherwise.

\subsection{Lagrangian description\label{sub:Lagrangian-description}}

Replacing $\boldsymbol{v}$ by $\partial_{\tau}\boldsymbol{X}$ in
Eq.~\eqref{eqn:intro}, we can make the action in the Lagrangian
description. The conservation laws of mass and entropy are given by
\begin{equation}
\rho J=\rho_{{\rm init}}\ {\rm and}\ s=s_{{\rm init}}\ ,\label{eq:original constraints}\end{equation}
 where $\rho_{{\rm init}}\ {\rm and}\ s_{{\rm init}}$ denote initial
values of $\rho$ and $s$, respectively. Thus, the action can be
defined as \begin{equation}
S_{{\rm L}}[\rho,s,\boldsymbol{X},\kappa,\lambda]\equiv\int_{t_{{\rm init}}}^{t_{{\rm fin}}}\!\!\!\!\!\! d\tau\!\int_{V}\!\! d^{3}\boldsymbol{a}\ \left\{ J{\cal L}(\rho,s,\partial_{\tau}\boldsymbol{X})\!+\! K(\rho J-\rho_{{\rm init}})\!+\!\rho\Lambda J(s-s_{{\rm init}})\right\} \ ,\label{eq:original action}\end{equation}
 where $K$ and $\Lambda$ are undetermined multipliers introduced
to keep the constraints (Eq.~\eqref{eq:original constraints}). In
this variational calculus, both ends of a pathline are fixed, \textit{\emph{i.e.}}\emph{,}
\begin{equation}
\delta X_{i}(\boldsymbol{a},t_{{\rm init}})=\delta X_{i}(\boldsymbol{a},t_{{\rm fin}})=0\ ,\label{eq:boundaries for X}\end{equation}
 where $\delta$ indicates an infinitesimal variation. Let us write
$\boldsymbol{n}$ for the unit normal vector of $\partial V$ directed
outside. The slip boundary condition, $n_{i}\partial_{\tau}X_{i}(\boldsymbol{a},\tau)=0\ {\rm if}\ \boldsymbol{a}\in\partial V,$
means that a fluid particle that is initially in contact with $\partial V$
remains in contact with $\partial V$ although it can slip along the
boundary. Thus, we have \begin{equation}
n_{i}\delta X_{i}(\boldsymbol{a},\tau)=0\ \ {\rm if}\ \ \boldsymbol{a}\in\partial V\ .\label{eq:boundary of space}\end{equation}

The stationary condition of Eq.~\eqref{eq:original action} with
respect to $K$, $\Lambda$, $\rho$, $s$, and $X_{i}$ is respectively
given by the two equations of Eq.~\eqref{eq:original constraints},\begin{eqnarray}
K & = & -\frac{1}{2}\left(\partial_{\tau}X_{i}\right)^{2}+h\ ,\label{eqn:rho1}\\
\Lambda & = & T\ ,\label{eqn:esu}\end{eqnarray}
 and \begin{equation}
\rho J\frac{\partial^{2}}{\partial\tau^{2}}X_{i}=-\frac{\partial}{\partial a_{j}}\left\{ \rho\left(\frac{1}{2}\left(\partial_{\tau}X_{k}\right)^{2}-\epsilon+K\right)\right\} \frac{\partial J}{\partial(\partial X_{i}/\partial a_{j})}\ ,\label{eqn:varilag}\end{equation}
 where $h$ is enthalpy defined as $h\equiv\epsilon+p/\rho$. The
undermined multipliers $K\ {\rm and}\ \Lambda$ are related to the
physical quantities by means of Eqs.~\eqref{eqn:rho1} and \eqref{eqn:esu}.
Note that surface integral terms, appearing when we applied the integration
by parts in calculating Eq.~\eqref{eqn:varilag}, vanish because
of the boundary conditions (Eq.~\eqref{eq:boundary of space}).

From Eqs.~\eqref{eqn:rho1}--\eqref{eqn:varilag}, we successfully
obtain the Euler equation in the Lagrangian description, \begin{equation}
\rho J\frac{\partial^{2}X_{i}}{\partial\tau^{2}}=-\frac{\partial p}{\partial a_{j}}\frac{\partial J}{\partial(\partial X_{i}/\partial a_{j})}\ .\label{eqn:ellag}\end{equation}
 The above is equivalent to the Euler equation in the Eulerian description,
which is given by Eq.~\eqref{eqn:euler} below. We can show the equivalence
by multiplying Eq.~\eqref{eqn:ellag} by $J^{-1}(\boldsymbol{x},t)$,
and replacing $\partial^{2}X_{i}/\partial\tau^{2}$ and $J^{-1}(\partial J/\partial(\partial X_{i}/\partial a_{j}))$
by $(\partial_{t}+\boldsymbol{v}\cdot\nabla)v_{i}$ and $\partial a_{j}/\partial x_{i}$,
respectively \cite{Bennett:}. As shown in Eq.~\eqref{eq:boundaries for X},
we fix both ends of a pathline. The Euler equation Eq.~\eqref{eqn:ellag}
is a second-order differential equation with respect to time $\tau$,
and its solution has two constants of integration, which are determined
by the fixed ends.

\subsection{Eulerian description\label{sub:Eulerian-description}}

In the Eulerian description, $\rho$, $s$, and $\boldsymbol{v}$
are regarded as functions of $\boldsymbol{x}=(x_{1},x_{2},x_{3})$
and $t$, and we need not replace $v_{i}$ by $\partial_{\tau}X_{i}$
in Eq.~\eqref{eqn:intro}. The boundary condition Eq.~\eqref{eq:boundary of space}
gives \begin{equation}
n_{i}v_{i}(\boldsymbol{x},t)=0\ \ {\rm if}\ \ \boldsymbol{x}\in\partial V\ ,\label{eq:boundary of v in the Eulerian}\end{equation}
 while the conservation laws of mass and entropy Eq.~\eqref{eq:original constraints}
can be rewritten into \begin{eqnarray}
D_{t}(\rho*1)=0\ , & \ {\rm i.e.},\  & \partial_{t}\rho=-\nabla\cdot(\rho\boldsymbol{v})\ ,\label{eq:conservation law of rho}\\
D_{t}s=0\ , & \ {\rm i.e.},\  & \partial_{t}s=-\boldsymbol{v}\cdot\nabla s\ ,\label{eq:conservation law of entropy}\end{eqnarray}
 where $D_{t}$ and $*1$ denote the Lagrangian time derivative equivalent
to $\partial_{\tau}$ and volume element equivalent to $J$, respectively.
See Appendix \ref{sec:The-Lagrangian-time} for the details. If we
follow a straightforward way from Eq.~\eqref{eq:original action},
the action in the Eulerian description should be given by \begin{equation}
S_{{\rm E}}[\rho,s,\boldsymbol{v},\kappa,\lambda]\!\equiv\!\int_{t_{{\rm init}}}^{t_{{\rm fin}}}\!\!\!\! dt\int_{V}\!\!\! d^{3}\!\boldsymbol{x}\left\{ \!{\cal L}(\rho,s,\boldsymbol{v})\!-\!\kappa\!\left(\!\frac{\partial\rho}{\partial t}\!+\!\nabla\cdot(\rho\boldsymbol{v})\!\right)\!-\!\lambda\rho\!\left(\!\frac{\partial s}{\partial t}\!+\!\boldsymbol{v}\cdot\nabla s\right)\!\right\} \ ,\label{eq:palin_lagrangian_in_eulreian_coordinate}\end{equation}
 where $\kappa\ {\rm and\ \lambda}$ are undetermined multipliers.
% for Eqs.~\eqref{eq:conservation law of rho} and \eqref{eq:conservation 
%law of entropy}. 
Keeping \begin{eqnarray}
\delta\rho(\boldsymbol{x},t_{{\rm init}})=\delta\rho(\boldsymbol{x},t_{{\rm fin}})=0\ ,\label{eq:initial rho}\\
\delta s(\boldsymbol{x},t_{{\rm init}})=\delta s(\boldsymbol{x},t_{{\rm fin}})=0\ ,\label{eq:initial s}\end{eqnarray}
 we find the stationary conditions of Eq.~\eqref{eq:palin_lagrangian_in_eulreian_coordinate}
with respect to $\kappa$, $\lambda$, $\boldsymbol{v}$, $\rho$,
and $s$ to be given respectively by Eqs.~\eqref{eq:conservation law of rho}
and \eqref{eq:conservation law of entropy}, \begin{eqnarray}
\boldsymbol{v} & = & -\nabla\kappa+\lambda\nabla s\ ,\label{eq:v}\\
D_{t}\kappa & = & -\frac{1}{2}\boldsymbol{v}^{2}+h\ ,\label{eqn:rho in Eulerian coordinate}\end{eqnarray}
 and \begin{equation}
D_{t}\lambda=T\ .\label{eqn:esu in Eulerian coordinate}\end{equation}
 Note that surface integral terms appearing in the calculation vanish
because of Eqs.~\eqref{eq:boundary of v in the Eulerian}, \eqref{eq:initial rho},
and \eqref{eq:initial s}. As discussed later in \S\ref{sub:Pontryagin's-minimum-principle},
we have the same stationary conditions, Eqs.~\eqref{eq:conservation law of rho},
\eqref{eq:conservation law of entropy}, and \eqref{eq:v}--\eqref{eqn:esu in Eulerian coordinate},
even if we replace Eqs.~\eqref{eq:initial rho} and \eqref{eq:initial s}
by other restrictions of $\rho$ and $s$ at the initial and final
times. Comparing Eqs.~\eqref{eqn:rho1}~and~\eqref{eqn:esu} with
Eqs \eqref{eqn:rho in Eulerian coordinate} and \eqref{eqn:esu in Eulerian coordinate},
respectively, we find $K=D_{t}\kappa$ and $\Lambda=D_{t}\lambda$.
With the aid of Eq.~\eqref{eq:conservation law of entropy}, the
Lagrangian time derivative of Eq.~\eqref{eq:v} yields \begin{equation}
D_{t}\boldsymbol{v}=-\nabla D_{t}\kappa+(D_{t}\lambda)\nabla s\ .\label{eq:euler0}\end{equation}
 Substituting Eqs.~\eqref{eqn:rho in Eulerian coordinate} and \eqref{eqn:esu in Eulerian coordinate}
into Eq.~\eqref{eq:euler0}, we obtain the Euler equation \begin{equation}
\frac{\partial}{\partial t}\boldsymbol{v}+\frac{1}{2}\nabla\boldsymbol{v}^{2}-\boldsymbol{v}\times(\nabla\times\boldsymbol{v})=-\frac{\nabla p}{\rho}\ .\label{eqn:euler}\end{equation}
 From Eq.~\eqref{eq:v}, the vorticity is found to be given by \begin{equation}
\boldsymbol{\omega}\equiv\nabla\times\boldsymbol{v}=\nabla\lambda\times\nabla s\ .\label{eq.vorticity0}\end{equation}
 Although the derivation of the Euler equation appears successful,
the resultant vorticity Eq.~\eqref{eq.vorticity0} vanishes on the
isentropic condition $\nabla s=0$. It is the flaw mentioned in the
introduction.

\section{Fixing ends of a pathline\label{sub:Fixing-the-endpoints}}

Both ends of a pathline are fixed in the Lagrangian variational calculus,
as shown in Eq.~\eqref{eq:boundaries for X}. It is thus natural
that the ends are fixed in the Eulerian variational calculus, but
not in \S\ref{sub:Eulerian-description}.

Let $A_{i}\ (i=1,2,3)$ be three scalar fields so that a pathline
coincides with an intersection of hypersurfaces given by \begin{equation}
A_{i}(\boldsymbol{x},t)={\rm constant}\ \ {\rm for}\ \ i=1,2,3\ .\label{eq:A}\end{equation}
 The determinant of the Jacobian matrix \begin{equation}
\frac{\partial(A_{1},A_{2},A_{3})}{\partial(x_{1},x_{2},x_{3})}=(\nabla A_{1}\times\nabla A_{2})\cdot\nabla A_{3}\label{eq:jacobian2}\end{equation}
 gives a reciprocal of the volume element in the coordinates defined
in terms of $\boldsymbol{A}$. The mass conservation, the first equation
of Eq.~\eqref{eq:original constraints}, is also represented by \begin{equation}
\rho\left(\boldsymbol{a}(\boldsymbol{x},t),t_{{\rm init}}\right)(\nabla A_{1}\times\nabla A_{2})\cdot\nabla A_{3}=\rho(\boldsymbol{x},t)\ ,\label{eqn:masconA}\end{equation}
 where $\boldsymbol{A}$ has variables $\boldsymbol{x}\ {\rm and}\ t$.

We can take $\boldsymbol{A}$ for the Lagrangian coordinate $\boldsymbol{a}$.
Otherwise, we can consider $\boldsymbol{A}$ as an invertible function
of $\boldsymbol{a}$. The indefiniteness was also pointed out in Ref.~\citen{Kambe}.
Suppose, for example, we have \begin{equation}
A_{1}=a_{1}\ ,A_{2}=a_{2},\ {\rm and}\ A_{3}=a_{3}+f(a_{1},a_{2})\ ,\label{eq:tilde A}\end{equation}
 where $f(a_{1},a_{2})$ is a two-variable function of $a_{1}$ and
$a_{2}$. Because we have \begin{equation}
\nabla A_{3}=\nabla a_{3}+\frac{\partial f}{\partial a_{1}}\nabla a_{1}+\frac{\partial f}{\partial a_{2}}\nabla a_{2}\ ,\label{eq:normal A}\end{equation}
 we can define $\nabla A_{3}$ so that it is normal to $\nabla A_{1}(=\nabla a_{1})$
and $\nabla A_{1}(=\nabla a_{2})$ by tuning the function $f$. If
we take $\tilde{\boldsymbol{A}}$ to be an arbitrary invertible function
of $\boldsymbol{a}$, and substitute $\tilde{\boldsymbol{A}}$, instead
of $\boldsymbol{a}$, into Eqs.~\eqref{eq:tilde A} and \eqref{eq:normal A},
we can also make $\nabla A_{3}$ normal to $\nabla A_{1}(=\nabla\tilde{A_{1}})$
and $\nabla A_{2}(=\nabla\tilde{A_{2}})$. By using it, we can rewrite
Eq.~\eqref{eqn:masconA} into \begin{equation}
\nabla A_{3}=\frac{\rho(\boldsymbol{x},t)}{\rho({\boldsymbol{a}(\boldsymbol{x},t),t_{{\rm init}}})(\nabla{A}_{1}\times\nabla{A}_{2})}\ ,\label{eq:massconA2}\end{equation}
 where $A_{i}$ has variables $\boldsymbol{x}\ {\rm and}\ t$.

Let us impose\begin{equation}
\delta A_{\alpha}(\boldsymbol{x},t_{{\rm init}})=\delta A_{\alpha}(\boldsymbol{x},t_{{\rm fin}})=0,\ {\rm for}\ \alpha=1,2\ .\label{eq:boundaries for endpoints in the Eulerian}\end{equation}
 Using Eqs.~\eqref{eq:initial rho}, \eqref{eq:massconA2}, and \eqref{eq:boundaries for endpoints in the Eulerian},
we have\begin{equation}
\delta A_{3}(\boldsymbol{x},t_{{\rm init}})=\delta A_{3}(\boldsymbol{x},t_{{\rm fin}})=0\ .\label{eqn:kore}\end{equation}
 Thus, we can fix the ends of a pathline in the variational calculus
by imposing Eqs.~\eqref{eq:initial rho} and \eqref{eq:boundaries for endpoints in the Eulerian}.
Note that the values of $A_{1}$ and $A_{2}$ can be determined independent
of the mass density $\rho$.

Since $\boldsymbol{v}$ is tangent to the hypersurfaces, $A_{1}$
and $A_{2}$ must satisfy \begin{equation}
D_{t}A_{\alpha}=0\ ,\ {\it {\rm i.e.}},\ \ \frac{\partial}{\partial t}A_{\alpha}=-\boldsymbol{v}\cdot\nabla A_{\alpha}\ .\label{eq:fixing endpoints}\end{equation}
 Hence, in the Eulerian description, we should minimize the action
\begin{equation}
S_{l}[\rho,s,\boldsymbol{v},\boldsymbol{A},\kappa,\lambda,\boldsymbol{\beta}]\equiv S_{{\rm E}}[\rho,s,\boldsymbol{v},\kappa,\lambda]-\int_{t_{{\rm init}}}^{t_{{\rm fin}}}\!\!\! dt\int_{V}dx^{3}\ \sum_{\alpha=1}^{2}\rho\beta_{\alpha}D_{t}A_{\alpha}\ ,\label{eq:action in Eulerian coordinate}\end{equation}
 with Eqs.~\eqref{eq:initial rho}, \eqref{eq:initial s}, and \eqref{eq:boundaries for endpoints in the Eulerian}
kept. Here, $\beta_{\alpha}$ is the undetermined multiplier, and
we write $\boldsymbol{A}$ and $\boldsymbol{\beta}$ for $(A_{1},A_{2})$
and $(\beta_{1},\beta_{2})$ respectively. The stationary conditions
of Eq.~\eqref{eq:action in Eulerian coordinate} with respect to
$A_{\alpha}$ are given by \begin{equation}
D_{t}\beta_{\alpha}=0\ ,\ {\rm i.e.,}\ \ \frac{\partial}{\partial t}\beta_{\alpha}=-\boldsymbol{v}\cdot\nabla\beta_{\alpha}\ .\label{eq:beta1}\end{equation}
 Here, surface integration terms appearing in the integration by parts
vanish because of Eqs.~\eqref{eq:boundary of v in the Eulerian}
and \eqref{eq:boundaries for endpoints in the Eulerian}. The stationary
conditions with respect to $\boldsymbol{v}\ {\rm and}\ \rho$ are
respectively given by \begin{equation}
\boldsymbol{v}=-\nabla\kappa+\lambda\nabla s+\sum_{\alpha=1}^{2}\beta_{\alpha}\nabla A_{\alpha}\ ,\label{eq:velocity obtained by Lin}\end{equation}
 and \begin{equation}
D_{t}\kappa=-\frac{1}{2}\boldsymbol{v}^{2}+h+\sum_{\alpha=1}^{2}\beta_{\alpha}D_{t}A_{\alpha}\ ,\label{eq:rho enthalpy}\end{equation}
 which equals Eq.~\eqref{eqn:rho in Eulerian coordinate} because
of Eq.~\eqref{eq:fixing endpoints}. The other stationary conditions
with respect to $\kappa$, $\lambda$, $\rho$, and $s$ are respectively
given by Eqs.~\eqref{eq:conservation law of rho}, \eqref{eq:conservation law of entropy},
\eqref{eqn:rho in Eulerian coordinate}, and \eqref{eqn:esu in Eulerian coordinate}.
We obtain the Euler equation Eq.~\eqref{eqn:euler} from Eq.~\eqref{eq:velocity obtained by Lin}
in the same way as we used in the preceding section. Values of $\beta_{1}\ {\rm and}\ \beta_{2}$
are determined by the fixed ends, as discussed later in $\S$\ref{sub:Pontryagin's-minimum-principle}.
From Eq.~\eqref{eq:velocity obtained by Lin}, the vorticity is found
to be given by\begin{equation}
\boldsymbol{\omega}=\nabla\lambda\times\nabla s+\sum_{\alpha=1}^{2}\nabla\beta_{\alpha}\times\nabla A_{\alpha}\ .\label{eq:vorticity in the Eulerian coordinates}\end{equation}
 This term $\sum_{\alpha=1}^{2}\nabla\beta_{\alpha}\times\nabla A_{\alpha}$
makes the flow rotational even on the isentropic condition $\nabla s=0$,
and the flaw mentioned in the introduction is removed.

\section{Hamiltonian formulation\label{sec:Hamiltonian-formulations-for} }

Hamiltonian formulations in analytical mechanics can be regarded as
a special case of more generalized formulation in control theory,
known as Pontryagin's minimum principle \cite{Pontryagin,Schulz}.
In \S\ref{subsec:pont}, we first give a brief review of this theory.
In \S\ref{sub:Pontryagin's-minimum-principle}, we apply it to the
dynamics of a perfect fluid to derive one of the canonical Hamiltonian
formulations with Clebsch potentials, which was previously derived
in a different way \cite{Y2,Salmon}.

\subsection{Brief review of the control theory\label{subsec:pont}}

Let $\boldsymbol{q}$ represent the state of a system to be controlled,
and $\boldsymbol{u}$ represent the input to this system, and we assume
the time evolution of the state to be given in terms of a function
of the state and input as \begin{equation}
\frac{d}{dt}\boldsymbol{q}=\boldsymbol{F}(\boldsymbol{q},\boldsymbol{u})\ .\label{eq:motion of states}\end{equation}
 The optimal input is determined so that a cost functional \begin{equation}
\int_{t_{{\rm init}}}^{t_{{\rm fin}}}dt\ L(\boldsymbol{q}(t),\boldsymbol{u}(t))\ ,\label{eq:value fanctional}\end{equation}
 where $L$ denotes a function, is minimized on condition that the
initial and final states are fixed, \textit{\emph{i.e.}}, \begin{equation}
\delta\boldsymbol{q}(t_{{\rm init}})=0\ ,\label{eq:initial condition of q}\end{equation}
 and \begin{equation}
\delta\boldsymbol{q}(t_{{\rm fin}})=0\ .\label{eq:transvasality 1}\end{equation}
 We define the undetermined multiplier, $\boldsymbol{p}$, which is
also called costate. The optimal input is obtained by minimizing \begin{eqnarray}
S[\boldsymbol{q},\boldsymbol{p},\boldsymbol{u}] & = & \int_{t_{{\rm init}}}^{t_{{\rm fin}}}dt\ \left\{ L(\boldsymbol{q},\boldsymbol{u})+\boldsymbol{p}\cdot\left(\frac{d}{dt}\boldsymbol{q}-\boldsymbol{F}(\boldsymbol{q},\boldsymbol{u})\right)\right\} \nonumber \\
 & = & \int_{t_{{\rm init}}}^{t_{{\rm fin}}}dt\ \left\{ -H(\boldsymbol{q},\boldsymbol{p},\boldsymbol{u})+\boldsymbol{p}\cdot\frac{d}{dt}\boldsymbol{q}\right\} \ ,\label{eq:pontryagin0}\end{eqnarray}
 where $H(\boldsymbol{q},\boldsymbol{p},\boldsymbol{u})$ is defined
as \begin{equation}
H(\boldsymbol{q},\boldsymbol{p},\boldsymbol{u})\equiv-L(\boldsymbol{q},\boldsymbol{u})+\boldsymbol{p}\cdot\boldsymbol{F}(\boldsymbol{q},\boldsymbol{u})\ .\label{eq:hamiltonian}\end{equation}

Let $\boldsymbol{u^{*}}(\boldsymbol{q},\boldsymbol{p})$ denote the
input minimizing Eq.~\eqref{eq:pontryagin0} on condition that $\boldsymbol{q}$
and $\boldsymbol{p}$ are given, and it is necessary for $\boldsymbol{u^{*}}$
to satisfy

\begin{equation}
\frac{\partial H(\boldsymbol{q},\boldsymbol{p},\boldsymbol{u}^{*})}{\partial u_{i}^{*}}=0\ .\label{eq:sup}\end{equation}
 Introducing \begin{equation}
H^{*}(\boldsymbol{q},\boldsymbol{p})\equiv H(\boldsymbol{q},\boldsymbol{p},\boldsymbol{u}^{*}(\boldsymbol{q},\boldsymbol{p}))\ ,\label{eq:optimal hamiltonian}\end{equation}
 we define the preoptimized action as\begin{equation}
S^{*}[\boldsymbol{q},\boldsymbol{p}]\equiv\int_{t_{{\rm init}}}^{t_{{\rm fin}}}dt\ \left\{ -H^{*}(\boldsymbol{q},\boldsymbol{p})+\boldsymbol{p}\cdot\frac{d}{dt}\boldsymbol{q}\right\} \ ,\label{eq:sup integral}\end{equation}
 which is not larger than $S[\boldsymbol{q},\boldsymbol{p},\boldsymbol{u}]$.
We can obtain the optimal input by solving the stationary conditions
of Eq.~\eqref{eq:sup integral} with respect to $\boldsymbol{p}$
and $\boldsymbol{q}$, which are given respectively by \begin{eqnarray}
\frac{dq_{i}}{dt} & = & \frac{\partial H^{*}(\boldsymbol{q},\boldsymbol{p})}{\partial p_{i}}\label{eq:canonical of state}\end{eqnarray}
 and \begin{eqnarray}
\frac{dp_{i}}{dt} & = & -\frac{\partial H^{*}(\boldsymbol{q},\boldsymbol{p})}{\partial q_{i}}\ .\label{eq:canonical of costate}\end{eqnarray}
 The boundary conditions for these equations are given by the initial
and final states, which are assumed to be fixed in Eqs.~\eqref{eq:initial condition of q}
and \eqref{eq:transvasality 1}. If we impose not Eq.~\eqref{eq:transvasality 1}
but a boundary condition \begin{equation}
\boldsymbol{p}(t_{{\rm fin}})=0\ ,\label{eq:transvasality 2}\end{equation}
 on the stationary conditions of Eq.~\eqref{eq:sup integral}, we
can derive Eqs.~\eqref{eq:canonical of state} and \eqref{eq:canonical of costate}.
As discussed in Ref.~\citen{Schulz}, we can have more general conditions
for the initial and final states. These conditions are called the
transversality conditions in control theory.

Let us assume $\boldsymbol{q}$ to represent the position of a material
particle, and define $\boldsymbol{F}$ as \begin{equation}
\boldsymbol{F}(\boldsymbol{q},\boldsymbol{u})\equiv\boldsymbol{u}\ ,\label{eq:input v}\end{equation}
 and we find the formulation above to be equivalent to that of analytical
mechanics for the particle.

\subsection{An application for a perfect fluid \label{sub:Pontryagin's-minimum-principle}}

We can generalize the formulation in the preceding subsection to cases
where the state variable and input variable are functions of space
and time, respectively. Let us consider $\rho,s$, and $\boldsymbol{A}$
in \S\ref{sub:Fixing-the-endpoints} to represent the state, and
$\boldsymbol{v}$ to represent the input variable. We can identify
Eq.~\eqref{eq:motion of states} with a set of Eqs.~\eqref{eq:conservation law of rho},
\eqref{eq:conservation law of entropy}, and \eqref{eq:fixing endpoints}
by defining $\boldsymbol{q}\equiv(\rho,s,\boldsymbol{A})$ and $\boldsymbol{u}\equiv\boldsymbol{v}$.
The costate is defined as $\boldsymbol{p}\equiv(-\kappa,-\rho\lambda,-\rho\boldsymbol{\beta})$.
The variations at the initial and final times satisfy Eqs.~\eqref{eq:initial rho},
\eqref{eq:initial s}, and \eqref{eq:boundaries for endpoints in the Eulerian}.
The function corresponding to Eq.~\eqref{eq:hamiltonian} is given
by\begin{equation}
{\cal H}(\boldsymbol{q},\boldsymbol{p},\boldsymbol{u})\equiv-{\cal L}+\kappa\nabla\cdot(\rho\boldsymbol{v})+\rho\lambda\boldsymbol{v}\cdot\nabla s+\sum_{\alpha=1}^{2}\rho\beta_{\alpha}\nabla A_{\alpha}\ .\label{eq:hamiltonian0}\end{equation}
 Using Eqs.~\eqref{eqn:intro} and \eqref{eq:boundary of v in the Eulerian},
we find \begin{equation}
\int_{V}\!\!\! d^{3}\!\boldsymbol{x}\ {\cal H}(\boldsymbol{q},\boldsymbol{p},\boldsymbol{u})=\int_{V}\!\!\! d^{3}\!\boldsymbol{x}\ \left\{ {\cal H}^{*}(\boldsymbol{q},\boldsymbol{p})-\frac{\rho}{2}(\boldsymbol{v}+\nabla\kappa-\lambda\nabla s-\sum_{\alpha=1}^{2}\beta_{\alpha}\nabla A_{\alpha})^{2}\right\} \ ,\end{equation}
 where we define \begin{equation}
{\cal H}^{*}(\boldsymbol{q},\boldsymbol{p})\equiv\rho\left\{ \epsilon(\rho,s)+\frac{1}{2}(-\nabla\kappa+\lambda\nabla s+\sum_{\alpha=1}^{2}\beta_{\alpha}\nabla A_{\alpha})^{2}\right\} \ .\label{eq:Pontryagin-1}\end{equation}
 Thus, the velocity field given by Eq.~\eqref{eq:velocity obtained by Lin}
turns out to minimize the action on condition that the state and costate
are given. We find that Eq.~\eqref{eq:action in Eulerian coordinate}
satisfies \begin{eqnarray}
S_{l}[\boldsymbol{q},\boldsymbol{p},\boldsymbol{u}] & = & \int_{t_{{\rm init}}}^{t_{{\rm fin}}}\!\!\! dt\int_{V}\!\!\! d^{3}\!\boldsymbol{x}\ \left\{ -{\cal H}(\boldsymbol{q},\boldsymbol{p},\boldsymbol{u})+\boldsymbol{p}\cdot\frac{d\boldsymbol{q}}{dt}\right\} \nonumber \\
 & \geq & \int_{t_{{\rm init}}}^{t_{{\rm fin}}}\!\!\! dt\int_{V}\!\!\! d^{3}\!\boldsymbol{x}\ \left\{ -{\cal H}^{*}(\boldsymbol{q},\boldsymbol{p})+\boldsymbol{p}\cdot\frac{d\boldsymbol{q}}{dt}\right\} \ ,\end{eqnarray}
 which is the preoptimized action, $S_{l}^{*}[\boldsymbol{q},\boldsymbol{p}]$.
Here, we have \begin{equation}
\boldsymbol{p}\cdot\frac{d\boldsymbol{q}}{dt}=-\kappa\frac{\partial\rho}{\partial t}-\rho\lambda\frac{\partial s}{\partial t}-\sum_{\alpha=1}^{2}\rho\beta_{\alpha}\frac{\partial A_{\alpha}}{\partial t}\ .\label{eq:}\end{equation}
 The stationary condition of the preoptimized action gives a set of
Eqs.~\eqref{eq:conservation law of rho}, \eqref{eq:conservation law of entropy},
and \eqref{eq:fixing endpoints}, and a set of Eqs.~\eqref{eqn:rho in Eulerian coordinate},
\eqref{eqn:esu in Eulerian coordinate}, and \eqref{eq:beta1}. These
sets can be rewritten respectively into \begin{eqnarray}
\frac{\partial q_{i}(\boldsymbol{x},t)}{\partial t} & = & \frac{\partial{\cal H}^{*}(\boldsymbol{q},\boldsymbol{p})}{\partial p_{i}}-\sum_{j=1}^{3}\frac{\partial}{\partial x_{j}}\frac{\partial{\cal H}^{*}(\boldsymbol{q},\boldsymbol{p})}{\partial(\partial p_{i}/\partial x_{j})}\label{eq:canonical of state2}\end{eqnarray}
 and \begin{eqnarray}
\frac{\partial p_{i}(\boldsymbol{x},t)}{\partial t} & = & -\frac{\partial{\cal H}^{*}(\boldsymbol{q},\boldsymbol{p})}{\partial q_{i}}+\sum_{j=1}^{3}\frac{\partial}{\partial x_{j}}\frac{\partial{\cal H}^{*}(\boldsymbol{q},\boldsymbol{p})}{\partial(\partial q_{i}/\partial x_{j})}\ ,\label{eq:canonical of costate2}\end{eqnarray}
 for $i=1,2,3,4$, which correspond to Eqs.~\eqref{eq:canonical of state}
and \eqref{eq:canonical of costate}, respectively. We can rewrite
Eqs.~\eqref{eq:canonical of state2} and \eqref{eq:canonical of costate2}
respectively into \begin{eqnarray}
q_{i}(\boldsymbol{x},t_{1}) & = & q_{i}(\boldsymbol{x},t_{{\rm init}})+\int_{t_{{\rm init}}}^{t_{1}}dt\ \left\{ \frac{\partial{\cal H}^{*}(\boldsymbol{q},\boldsymbol{p})}{\partial p_{i}}-\sum_{j=1}^{3}\frac{\partial}{\partial x_{j}}\frac{\partial{\cal H}^{*}(\boldsymbol{q},\boldsymbol{p})}{\partial(\partial p_{i}/\partial x_{j})}\right\} \ ,\label{eq:solution of states}\\
p_{i}(\boldsymbol{x},t_{1}) & = & p_{i}(\boldsymbol{x},t_{{\rm init}})+\int_{t_{{\rm init}}}^{t_{1}}dt\ \left\{ -\frac{\partial{\cal H}^{*}(\boldsymbol{q},\boldsymbol{p})}{\partial q_{i}}+\sum_{j=1}^{3}\frac{\partial}{\partial x_{j}}\frac{\partial{\cal H}^{*}(\boldsymbol{q},\boldsymbol{p})}{\partial(\partial q_{i}/\partial x_{j})}\right\} \ .\label{eq:solution of costates}\end{eqnarray}
 In \S\ref{sub:Fixing-the-endpoints}, we fixed the initial and final
states by imposing Eqs.~\eqref{eq:initial rho}, \eqref{eq:initial s},
and \eqref{eq:boundaries for endpoints in the Eulerian}. These restrictions
can be rewritten into Eqs.~\eqref{eq:initial condition of q} and
\eqref{eq:transvasality 1}. The initial value of the costate $\boldsymbol{p}(\boldsymbol{x},t_{{\rm init}})$,
\textit{\emph{i.e.}}\emph{,} $\kappa(\boldsymbol{x},t_{{\rm init}}),\lambda(\boldsymbol{x},t_{{\rm init}}),$
and $\boldsymbol{\beta}(\boldsymbol{x},t_{{\rm init}})$, should be
determined so that the given values of $\boldsymbol{q}(\boldsymbol{x},t_{{\rm init}})$
and $\boldsymbol{q}(\boldsymbol{x},t_{{\rm fin}})$ are compatible
with Eqs.~\eqref{eq:solution of states} and \eqref{eq:solution of costates}.

We can relax the restrictions in the variational calculus. First,
suppose that we do not impose Eq.~\eqref{eq:boundaries for endpoints in the Eulerian}
for the final time. As discussed in relation to Eq.~\eqref{eq:transvasality 2},
we have $\boldsymbol{\beta}(\boldsymbol{x},t_{{\rm fin}})=\boldsymbol{0}$.
By substituting it into Eq.~\eqref{eq:beta1}, we obtain\begin{equation}
\boldsymbol{\beta}(\boldsymbol{x},t)=\boldsymbol{0}\ .\label{eqn:anytime beta}\end{equation}
 Thus, the velocity field Eq.~\eqref{eq:velocity obtained by Lin}
becomes irrotational on the isentropic condition, like the velocity
given by Eq.~\eqref{eq:v}. Next, suppose we do not impose any of
Eqs.~\eqref{eq:initial rho}, \eqref{eq:initial s}, and \eqref{eq:boundaries for endpoints in the Eulerian}
for the final time, and we have Eq.~\eqref{eqn:anytime beta} and
\begin{equation}
\kappa(\boldsymbol{x},t_{{\rm fin}})=\lambda(\boldsymbol{x},t_{{\rm fin}})=0\ .\label{eq:final kappa lambda}\end{equation}
 If we can solve Eqs.~\eqref{eq:solution of states} and \eqref{eq:solution of costates}
with their boundary conditions above, the solution uniquely determines
$\rho,s,\boldsymbol{A},\kappa,\lambda,\ {\rm and}\ \boldsymbol{\beta}$,
and thus $\boldsymbol{v}$, for any $\boldsymbol{x}$ and $t$ considered.

\section{Discussion \label{sec:Discussion}}

Various actions for a perfect fluid in the Eulerian description were
proposed in previous works so that a rotational isentropic flow can
be derived. Bateman\cite{Bateman} proposed an action with additional
scalar field $\zeta$ and $\eta$ as \begin{equation}
S_{b}[\rho,s,\boldsymbol{v},\kappa,\lambda,\eta,\zeta]\equiv S_{{\rm E}}[\rho,s,\boldsymbol{v},\kappa,\lambda]-\int dtdx^{3}\ \rho\zeta D_{t}\eta\ ,\label{eq:non-relativistic_lagrangian_in_eulreian_coordinate}\end{equation}
 and derived\begin{equation}
\boldsymbol{v}=-\nabla\kappa+\lambda\nabla s+\zeta\nabla\eta\ .\label{eq:velocity obtained by Bateman}\end{equation}
 On the isentropic condition, the velocity field is given by \begin{equation}
\boldsymbol{v}=-\nabla\kappa+\zeta\nabla\eta\ .\label{eq:clebsch}\end{equation}
 The additional field $\zeta$ and $\eta$ are sometimes called Clebsch
potentials because, apart from the variational principle, Clebsch
\cite{Clebsch,lamb} had shown that a velocity field of a perfect
fluid is given by Eq.~\eqref{eq:clebsch}, provided that the vorticity
can be put in the form\begin{equation}
\boldsymbol{\omega}=\nabla\zeta\times\nabla\eta\ .\label{eq:vorticity by Clebsch}\end{equation}

Lin \cite{Lin} introduced more scalar fields to identify a fluid
particle, mentioned the conservation law of identity without sufficient
explanation of physical meanings of the identities and their conservation,
and derived \begin{equation}
\boldsymbol{v}=-\nabla\kappa+\lambda\nabla s+\sum_{i=1}^{3}\beta_{i}\nabla A_{i}\ ,\label{eqn:threeA}\end{equation}
 which can be rotational on the isentropic condition. He considered
$A_{i}$ to be a function of the original position of the fluid particle.
Later, Selinger and Whitham \cite{SELINGER} mentioned, {}``Lin's
device still remains somewhat mysterious from a strictly mathematical
viewpoint, but necessary for it seems to be firmly established as
we proceed,'' and also commented, {}``But it was noted by Clebsch
that any flow velocity can be represented by'' Eq.~\eqref{eq:clebsch}.
This statement would be misleading because Eq.~\eqref{eq:clebsch}
is valid if and only if Eq.~\eqref{eq:vorticity by Clebsch} holds
\cite{lamb}. Schutz \cite{Schutz} generalized Bateman's variational
principle for a relativistic perfect fluid, and mentioned, {}``The
existence of an Eulerian variational principle may be a beginning.
What is needed, I believe, is a Hamiltonian principle in a minimum
number of variables. The present action principle seems to have 'too
many' free variables''. Yoshida\cite{Y2} pointed out that an arbitrary
velocity field can be given by Eq.~\eqref{eq:velocity obtained by Lin},
not Eq.~\eqref{eq:clebsch}, in a way different from the way we used
in \S \ref{sub:Fixing-the-endpoints}. There, we first introduced
$A_{i}$ $(i=1,2,3)$, which can be the Lagrangian coordinates, but
can be otherwise. Using the mass conservation Eq.~\eqref{eq:massconA2},
we next showed that we can fix both ends of a pathline without using
$A_{3}$.

Clebsch potentials have been considered to be related with the Lagrangian
coordinates \cite{Bennett:,Lin,SELINGER,Kambe,Yoshida,Holm,holm2}.
Selinger and Whitman \cite{SELINGER} mentioned that Clebsch potentials
satisfy Eq.~\eqref{eq:fixing endpoints}. Kambe \cite{Kambe} considered
Eq.~\eqref{eqn:threeA} to give the transformation between the Lagrangian
and Eulerian coordinates, and mentioned the indefiniteness of Clebsch
potentials. Yoshida \cite{Yoshida} claimed that Clebsch potentials
provide a one-to-one map between the Lagrangian densities in the Lagrangian
and Eulerian descriptions.

Although Clebsch potentials have been considered to label a fluid
particle, why the label is required in the Eulerian variational calculus
has not been clear in previous studies. We can fix both ends of a
path line in the Eulerian variational calculus by imposing Eq.~\eqref{eq:boundaries for endpoints in the Eulerian}
on the Clebsch potentials. In doing so, we can keep the same conditions
in the variational calculus, as in the Lagrangian description. Thus,
we can start with the same Lagrangian density to obtain the same velocity
field, which can be irrotational on the isentropic condition, as in
the Lagrangian description. As discussed in $\S$\ref{sub:Pontryagin's-minimum-principle},
if we do not impose Eq.~\eqref{eq:boundaries for endpoints in the Eulerian},
i.e.\emph{,} do not fix both ends of a path line, the resultant Eulerian
variational calculus has no corresponding Lagrangian variational calculus,
in which both ends of a path line are always fixed by Eq.~\eqref{eq:boundaries for X}.

As referred to in Ref.~\citen{Kambe}, Lin's variational principle
yields \begin{equation}
D_{t}\beta_{i}=0,\ \ {\rm for}\ \ i=1,2,3\label{eq:kambe}\end{equation}
 which means that $\beta_{i}$ is a function of the Lagrangian coordinates,
i.e., Clebsch potentials $A_{i}$ $(i=1,2,3)$, and the vorticity
on the isentropic condition is given by \begin{equation}
\boldsymbol{\omega}(A_{1},A_{2},A_{3})=\sum_{i,j=1}^{3}\frac{\partial\beta_{i}(\boldsymbol{A})}{\partial A_{j}}\nabla A_{j}\times\nabla A_{i}\ .\label{eq:vorticity in LD1}\end{equation}
 We showed that $A_{3}$ is a function of $\rho,A_{1},\ {\rm and}\ A_{2}$
in Eq.~\eqref{eq:massconA2}. Thus, the values of $\beta_{1}$ and
$\beta_{2}$ are given by functions of $\rho,A_{1},\ {\rm and}\ A_{2}$
because of Eq.~\eqref{eq:beta1}. Hence, the vorticity on the isentropic
condition can be rewritten into \begin{equation}
\boldsymbol{\omega}(A_{1},A_{2},\rho)=\left(\frac{\partial\beta_{2}}{\partial A_{1}}-\frac{\partial\beta_{1}}{\partial A_{2}}\right)\nabla A_{1}\times\nabla A_{2}+\sum_{\alpha=1}^{2}\frac{\partial\beta_{\alpha}}{\partial\rho}\nabla\rho\times\nabla A_{\alpha}\ .\label{eq:vorticity in LD-2}\end{equation}

Hamiltonian formulations for a perfect fluid also have been studied
by some authors \cite{Yoshida,Y2,Holm,cendra,holm2,Arnold,Morrison,Salmon}.
Arnold and Khesin \cite{Arnold} studied a Hamiltonian formulation
of an incompressible isentropic flow, and Morrison and Greene \cite{Morrison}
discovered an extremely complicated Eulerian Poisson bracket for an
isentropic perfect fluid. Holm and Kupershmidt \cite{Holm} modified
the Hamiltonian formulations \cite{Arnold,Morrison} by means of Clebsch
potentials. As pointed out in Ref.~\citen{Yoshida}, Casimir invariants,
such as helicity, constrain the dynamics of a perfect fluid in these
noncanonical Hamiltonian formulations \cite{Holm,cendra,holm2,Arnold,Morrison}.
Note that the helicity is conserved only for a barotropic perfect
fluid and an isentropic perfect fluid. On the other hand, as discussed
in Appendix \ref{sub:Poisson-bracket}, there is no Casimir invariant
in a canonical Hamiltonian formulation. We obtain the same canonical
Hamiltonian formulation as derived in Ref.~\citen{Y2} by regarding
the velocity as the input variable.

We have assumed fluid dynamics in a fixed container. This assumption
can be relaxed; our discussion remains valid when we consider the
dynamics of a perfect fluid in an infinitely large spacial region
and assume that the mass density tends to become zero far from a certain
point. Our discussion can be applied to a relativistic prefect fluid,
where we should start with the Lagrangian density slightly different
from Eq.~\eqref{eqn:intro}. We show the details in Appendix \ref{sec:A-variational-principle}
for completeness of our study.

\section*{Acknowledgements}

We thank T. Kambe, who turned our interest to this problem. H. F.
also thanks Z. Yoshida, Y. Hattori, T. K. Nakamura, A. Hosoya, and
I. Ojima for valuable discussions. The work by H. F. was financially
supported by the KLL Research Grant for Ph.D. Program. Part of the
work by Y. F. was financially supported by the Keio Gijuku Academic
Development Funds.

\appendix
%dummy comment inserted by tex2lyx to ensure that this paragraph is not empty

\section{Lagrangian Time Derivative \label{sec:The-Lagrangian-time}}

The Lagrangian time derivative $\partial_{\tau}\equiv D_{t}$ is given
by $\partial_{t}+\boldsymbol{v}\cdot\nabla\ $ for a scalar, $\partial_{t}+\nabla(\boldsymbol{v}\cdot\ )-\boldsymbol{v}\times\nabla\times\ $
for a cotangent vector, $\partial_{t}-\nabla\times(\boldsymbol{v}\times\ )+\boldsymbol{v}(\nabla\cdot\ )$
for an axial vector, and $\partial_{t}+\nabla\cdot(\boldsymbol{v}\ )$
for the volume. These expressions are unified as \begin{equation}
D_{t}\equiv\partial_{t}+L_{\boldsymbol{v}}\ ,\label{eq:-3}\end{equation}
 where $L_{\boldsymbol{v}}$ denotes the Lie derivative along the
vector field $\boldsymbol{v}$. Note that $D_{t}$ is commutative
with exterior derivative $d$, and thus gradient $\nabla\cdot\ $
and rotation $\nabla\times\ $. Because $\rho$ and $s$ are 0-forms,
and the volume element $*1$ is a 3-form \cite{Schutzgeo}, the conservation
laws of mass and entropy are respectively found to be given by\begin{equation}
D_{t}(\rho*1)=0\ \ {\rm and}\ \ D_{t}s=0\ .\label{eq:-2}\end{equation}

\section{Canonical Poisson Bracket\label{sub:Poisson-bracket}}

Let $f$ and $g$ denote functions of $\boldsymbol{q}$ and $\boldsymbol{p}$,
which are $N$-dimensional vectors, and we define their Poisson bracket
as \begin{equation}
\{f,g\}=\sum_{i=1}^{N}\left[\frac{\delta f}{\delta q_{i}}\frac{\delta g}{\delta p_{i}}-\frac{\delta f}{\delta p_{i}}\frac{\delta g}{\delta q_{i}}\right]\ ,\label{eq:poison bracket}\end{equation}
 where we define \begin{eqnarray}
\frac{\delta}{\delta q_{i}} & \equiv & \frac{\partial}{\partial q_{i}}-\sum_{j=1}^{3}\frac{\partial}{\partial x_{j}}\frac{\partial}{\partial(\partial q_{i}/\partial x_{j})}\ ,\\
\frac{\delta}{\delta p_{i}} & \equiv & \frac{\partial}{\partial p_{i}}-\sum_{j=1}^{3}\frac{\partial}{\partial x_{j}}\frac{\partial}{\partial(\partial p_{i}/\partial x_{j})}\ .\end{eqnarray}
 We can rewrite Eqs.~\eqref{eq:canonical of state} and \eqref{eq:canonical of costate}
into \begin{eqnarray}
\frac{d}{dt}\boldsymbol{q} & = & \{\boldsymbol{q},{\cal H}^{*}(\boldsymbol{q},\boldsymbol{p})\}\ ,\\
\frac{d}{dt}\boldsymbol{p} & = & \{\boldsymbol{p},{\cal H}^{*}(\boldsymbol{q},\boldsymbol{p})\}\ ,\end{eqnarray}
 respectively. This Hamiltonian formulation is canonical because the
associated Poisson bracket can be rewritten by means of the symplectic
matrix, \textit{\emph{i.e.,}} \begin{equation}
\{f,g\}=\left(\frac{\delta f}{\delta q_{i}},\frac{\delta f}{\delta p_{i}}\right)\left(\begin{array}{cc}
0 & I\\
-I & 0\end{array}\right)\left(\frac{\delta g}{\delta q_{i}},\frac{\delta g}{\delta p_{i}}\right)^{t}\ ,\label{eq:poison bracket1}\end{equation}
 where $I$ and the superscript $^{t}$ are the $N\times N$ unit
matrix and the transposition, respectively. A Casimir invariant $C$
is defined so that $\{F,C\}=0$ for any $F$, and is a conserved quantity
%\ .\label{eq:casimir}\end{equation}
since ${\rm d}C/{\rm d}t=\{H,C\}=0$. Thus, if $C$ is a Casimir invariant,
we have \begin{equation}
\left(\begin{array}{cc}
0 & I\\
-I & 0\end{array}\right)\left(\frac{\delta C}{\delta q_{i}},\frac{\delta C}{\delta p_{i}}\right)^{t}\ =0\ ,\label{eq:kernel}\end{equation}
 which tells that $C$ does not depend on any of $\boldsymbol{q}$
and $\boldsymbol{p}$, \textit{\emph{i.e.}}\emph{,} that $C$ is trivial.

Because Eqs.~\eqref{eq:canonical of state2} and \eqref{eq:canonical of costate2}
can be rewritten by means of the symplectic matrix, our Hamiltonian
formulation given in \S\ref{sub:Pontryagin's-minimum-principle}
is canonical.

\section{Formulation for a Relativistic Perfect Fluid\label{sec:A-variational-principle}}

Let us consider a relativistic perfect fluid in a four-dimensional
space-time region $\Omega$. We redefine $\rho$ and $s$ as the particle
number density and the entropy per particle in the local rest frame
of matter, respectively, and define $\boldsymbol{u}$ as the four-velocity
field. We assume that metric tensor $g(\ ,\ )$ is given by other
materials, and put the speed of light equal to unity. The normalization
of the four-velocity field $\boldsymbol{u}$ is given by \begin{equation}
g(\boldsymbol{u},\boldsymbol{u})+1=0\ .\label{eq:normalization}\end{equation}
 The constraints Eqs.~\eqref{eq:conservation law of rho}, \eqref{eq:conservation law of entropy},
and \eqref{eq:fixing endpoints} are respectively rewritten as\begin{equation}
L_{\boldsymbol{u}}(\rho*1)=0,\ L_{\boldsymbol{u}}s=0,\ \ {\rm and}\ \ L_{\boldsymbol{u}}A_{\alpha}=0,\ \ {\rm for}\ \alpha=1,2\ .\label{eq:hyper}\end{equation}
 The Lagrangian density for a relativistic perfect fluid is given
by\begin{equation}
{\cal L}(\rho,s)=-\rho\epsilon\ .\label{eq:Lagrangian density in relativity}\end{equation}
 Let $*1$ denote the four-dimensional volume element, and we can
modify the action Eq.~\eqref{eq:action in Eulerian coordinate} as
\begin{eqnarray}
S(\rho,s,\boldsymbol{A,}\kappa,\lambda,\boldsymbol{\beta},\boldsymbol{u},\gamma) & \!\!=\!\! & \int_{\Omega}\Bigl\{*1{\cal L}(\rho,s)+\gamma*1\left\{ g(\boldsymbol{u},\boldsymbol{u})+1\right\} \nonumber \\
 & \!\!\!\! & \!\!-\!\!\kappa L_{\boldsymbol{u}}(\rho*1)\!-\!\rho\lambda*1L_{\boldsymbol{u}}s\!-\!\!\sum_{\alpha=1}^{2}\rho\beta_{\alpha}*1L_{\boldsymbol{u}}A_{\alpha}\Bigr\}\ ,\label{eq:action with restraint conditions}\end{eqnarray}
 where $\gamma$ is an undetermined multiplier for the normalization.
We still have Eq.~\eqref{eq:boundary of v in the Eulerian}, and
impose restrictions Eqs.~\eqref{eq:initial rho}, \eqref{eq:initial s}, and \eqref{eq:boundaries for endpoints in the Eulerian} in the variational calculus. The
stationary conditions corresponding to Eqs.~\eqref{eqn:rho in Eulerian coordinate},
and \eqref{eqn:esu in Eulerian coordinate} are respectively given
by\begin{eqnarray}
 &  & -h+L_{\boldsymbol{u}}\kappa=0\ ,\label{eq:rho}\\
 &  & -T+L_{\boldsymbol{u}}\lambda=0\ ,\label{eq:s}\\
 &  & L_{\boldsymbol{u}}\beta_{\alpha}=0\ ,\label{eq:alpha}\\
 &  & 2\gamma g(\boldsymbol{u},\ )+d\kappa-\lambda ds-\sum_{\alpha=1}^{2}\beta_{\alpha}dA_{\alpha}=\boldsymbol{0}\ ,\label{eq:u}\end{eqnarray}
 where enthalpy $h$, pressure $p$, and temperature $T$ are defined
in the local rest frame of matter. From Eqs.~\eqref{eq:normalization},
\eqref{eq:hyper}, \eqref{eq:rho}--\eqref{eq:u}, we obtain $2\gamma=h$
and thus have the equation for the four-momentum field\begin{equation}
hg(\boldsymbol{u},\ )+d\kappa-\lambda ds-\sum_{i=1}^{2}\beta_{\alpha}dA_{\alpha}=\boldsymbol{0}\ ,\label{eq:four momentum}\end{equation}
 of which the Lie derivative gives the Euler equation\begin{equation}
L_{\boldsymbol{u}}\left\{ hg(\boldsymbol{u},\ )\right\} +dp/\rho=\boldsymbol{0}\ .\label{eq:relativistic Euler}\end{equation}

Defining \begin{equation}
\boldsymbol{v}\equiv\left(\frac{u_{1}}{u_{0}},\frac{u_{2}}{u_{0}},\frac{u_{3}}{u_{0}}\right)\ ,\end{equation}
 we have $u_{0}=1/\sqrt{1-\boldsymbol{v}^{2}}$ from Eq.~\eqref{eq:normalization}.
As in \S \ref{sub:Pontryagin's-minimum-principle}, we take $\bar{\rho}\equiv\rho u_{0},\ s,\ {\rm and}\ \boldsymbol{A}$
to be the state $\boldsymbol{q}$, take $-\kappa,-\rho\lambda\ {\rm and}\ -\rho\boldsymbol{\beta}$
to be the costate $\boldsymbol{p}$, and take $\boldsymbol{v}$ to
be the input. Because Eq.~\eqref{eq:hyper} gives \begin{eqnarray}
\frac{\partial}{\partial t}\bar{\rho} & = & -\nabla\cdot(\bar{\rho}\boldsymbol{v})\ ,\\
\frac{\partial}{\partial t}s & = & -\boldsymbol{v}\cdot\nabla s\ ,\\
\frac{\partial}{\partial t}A_{\alpha} & = & -\boldsymbol{v}\cdot\nabla A_{\alpha}\ ,\end{eqnarray}
 we can find the function corresponding to Eq.~\eqref{eq:hamiltonian0}
to be given by \begin{equation}
{\cal H}(\rho,s,\boldsymbol{A},\lambda,\kappa,\boldsymbol{\beta},\boldsymbol{v})=-{\cal L}(\bar{\rho}\sqrt{1-\boldsymbol{v}^{2}},s)+\kappa\nabla\cdot(\bar{\rho}\boldsymbol{v})+\bar{\rho}\lambda\boldsymbol{v}\cdot\nabla s+\sum_{i=\alpha}^{2}\bar{\rho}\beta_{\alpha}\boldsymbol{v}\cdot\nabla A_{\alpha}\ ,\end{equation}
 and the optimized input $\boldsymbol{v}^{*}(\boldsymbol{q},\boldsymbol{p})$
to satisfy Eq.~\eqref{eq:four momentum}.


\begin{thebibliography}{21}
\bibitem{Bennett:} A. Bennett, \textit{Lagrangian fluid dynamics}
(Cambridge Univ. Press, Cambridge, 2006), p. 32.

\bibitem{Bateman}H. Bateman, Proc. Roy. Soc. London. A \textbf{125}
(1929), 598; Scripta Math. \textbf{10} (1944), 51; \textit{Partial
differential equations} (Dover, New York, 1944), p. 164.

\bibitem{Clebsch}A. Clebsch, J. Reine Angew. Math. \textbf{56} (1859),
1.

\bibitem{lamb}H. Lamb, \textit{Hydrodynamics} (Cambridge Univ. Press,
Cambridge, 1932), p. 248.

\bibitem{Lin}C. C. Lin, in \textit{International School of Physics
Enrico Fermi (XXI)}, ed. G. Careri (Academic Press, New York, 1963),
p. 93.

\bibitem{SELINGER}R. L. Selinger and G. B. Whitham, Proc. Roy. Soc.
London. A \textbf{305} (1968), 1.

\bibitem{Schutz} B. F. Schutz, Jr, Phys. Rev. D \textbf{2} (1970),
2762; Phys. Rev. D \textbf{4} (1971), 3559.

\bibitem{Elze}H.-T. Elze, Y. Hama, T. Kodama, M. Makler and J. Rafelski,
J. Phys. G. \textbf{25} (1999), 1935.

\bibitem{Friedman}J. L. Friedman and J. R. Ipser, Philos. Trans.
R. Soc. London, Ser. A \textbf{340} (1992), 39.

\bibitem{Kambe}T. Kambe, Fluid Dyn. Res. \textbf{39} (2007), 98;
Fluid Dyn. Res. \textbf{40} (2008), 399; Physica D \textbf{237} (2008),
2067; \textit{Geometrical theory of dynamical systems and velocity
field}, Revised ed., (World Scientific, Singapore, 2010), p. 189.

\bibitem{Yoshida}Z. Yoshida,\textit{ Proc. Int. Symp.} \textit{Contemporary
Physics}, ed. J. Aslam, F. Hussain and Riazuddin (World Scientific,
Singapore, 2008), p. 125.

\bibitem{Y2} Z. Yoshida, J. Math. Phys. \textbf{50} (2009), 113101.

\bibitem{Holm}D. D. Holm and B. A. Kupershmidt, Physica D \textbf{7}
(1983), 330.

\bibitem{cendra}H. Cendra and J. E. Marsden, Physica D \textbf{27}
(1987), 63.

\bibitem{holm2}D. D. Holm, J. E. Marsden and T. S. Ratiu, Adv. Math.
\textbf{137} (1998), 1.

\bibitem{Arnold}V. I. Arnold and B. A. Khesin, \emph{Topological
methods in hydrodynamics} (Springer-Verlag, Berlin, 1997), p. 47.

\bibitem{Morrison}P. J. Morrison and J. M. Greene, Phys. Rev. Lett.
\textbf{45} (1980), 790; Phys. Rev. E \textbf{48} (1982), 569.

\bibitem{Salmon}R. Salmon, Ann. Rev. Fluid Mech. \textbf{20} (1988),
225.

\bibitem{Pontryagin}L. S. Pontryagin, V. G. Boltyanskii, R. V. Gamkrelidze
and E. F. Mischenko, \emph{The mathematical theory of optimal processes}
(Gordon \& Breach Science Pub, New York, 1987), p. 66.

\bibitem{Schulz}M. Schulz, \emph{Control theory in physics and other
fields of science} (Springer-Verlag, Berlin, 2006), p. 17.

\bibitem{Schutzgeo} B. F. Schutz, \textit{Geometrical methods of
mathematical physics} (Cambridge Univ. Press, Cambridge, 1980), p.
181. 
\end{thebibliography}
\end{document}